\documentclass[epj]{svjour}
\def\makeheadbox{{%
\hbox to0pt{\vbox{\baselineskip=10dd\hrule\hbox
to\hsize{\vrule\kern3pt\vbox{\kern3pt
\hbox{Eur. Phys. J. E \textbf{34} (2011) 115. DOI:~10.1140/epje/i2011-11115-7}
\kern3pt}\hfil\kern3pt\vrule}\hrule}%
\hss}}}

\usepackage[english]{babel}
\usepackage[latin1]{inputenc}
\usepackage{amsmath,bbm}
\usepackage{graphicx}
\usepackage{eulervm}
\newcommand{\bfe}{{\mathbf{e}}}
\newcommand{\bfk}{{\mathbf{k}}}

\newcommand{\bfq}{{\mathbf{q}}}

\newcommand{\bfu}{{\mathbf{u}}}
\newcommand{\bfv}{{\mathbf{v}}}

\newcommand{\hatu}{\hat{u}}
\newcommand{\hatbfu}{\hat{\bfu}}

\newcommand{\tilmu}{\tilde{\mu}}
\newcommand{\tillambda}{\tilde{\lambda}}
\newcommand{\diag}{\operatorname{diag}}
\newcommand{\grad}{\operatorname{grad}}
\renewcommand{\div}{\operatorname{div}}

\begin{document}
\title{Anisotropic elasticity  in confocal studies of colloidal crystals}
\author{Michael Schindler \and A. C. Maggs}
\institute{Laboratoire PCT, UMR Gulliver CNRS-ESPCI 7083, 10 rue
  Vauquelin, 75231 Paris Cedex 05}
\date{\today}

\abstract{
  We consider the theory of fluctuations of a colloidal solid observed
  in a confocal slice. For a cubic crystal we study the evolution of
  the projected elastic properties as a function of the anisotropy of
  the crystal using numerical methods based on the fast Fourier
  transform. In certain situations of high symmetry we find exact
  analytic results for the projected fluctuations.
}
\maketitle

\section{Introduction}

In a recent paper~\cite{claire} we simulated a large ensemble of hard
spheres packed in a face-centered cubic crystal in order to study the
fluctuations within a slice of the sample. This work was motivated by
the interpretation of the elastic properties found in experiments of
colloidal crystals~\cite{maret,maret2,andrea,science,kurchan,antinaPRL}. In particular we showed how to extract
information on the three-dimensional elastic constants from the
anomalous dispersion relationship that is observed within a single
slice. While we managed to perform an exact analytic calculation for
isotropic elastic media many experiments are performed on crystals,
rather than amorphous isotropic solids. The question then arises as to
how to best treat the anisotropy. We proposed using an averaging
procedure such as that of Fedorov~\cite{Norris06,Fedorov68} and for
the sample that we simulated the results were remarkably good:
Simulation and theory agreed to within a few percent. However, many
months of simulation are required to study a single sample and it is
difficult to perform a systematic study of the effect of anisotropy on
the projected properties.

In this paper we study the projected fluctuations within linear
elasticity, rather than by molecular dynamics simulation. By
formulating cubic elastic properties on a discrete grid we are able
to evaluate the anisotropic Green function within a projection plane
using fast Fourier transforms. This procedure allows us to study how
well our analytic approximation reproduces the exact projected
properties as a function of the plane orientation as well as of the
three elastic constants characterizing the full cubic system.

\section{Cubic Elasticity}

Let us now consider a three-dimensional cubic crystal of atoms,
indexed by~$i$, deformed by the displacement vector~$u_i$. We write
the elastic energy as a quadratic form in the spatial derivatives
$u_{ij}$ of this vector. This quadratic form, which respects the
symmetry of the crystal, can be expressed with the help of the
Christoffel matrix~\cite{Wallace70}
\begin{equation}
  \label{eq:christoffel}
  D_{ik}(\bfk) = \Bigl[\lambda\delta_{ij}\delta_{kl}
  + \mu(\delta_{ik}\delta_{jl} + \delta_{il}\delta_{jk})
  + \nu S_{ijkl}\Bigr] k_j k_l,
\end{equation}
with a reciprocal vector~$\bfk$, Lam\'e constants $\lambda$, $\mu$, and
anisotropy~$\nu$. The tensor $S=\sum_{p=1}^3 \bfe^{p} \bfe^{p}
\bfe^{p} \bfe^{p}$, with ${\bfe}^p$ unit vectors parallel to the
cubic axes of the crystal. Summation over indices occurring twice is
assumed. In a colloidal crystal, being under external pressure~$P$,
the constants are modified by non-linear
interactions~\cite{Wallace70}. The above constants are then related to
the elastic constants used in Voigt notation,
\begin{align*}
  C_{11}-P&=\lambda+ 2 \mu +\nu\\
  C_{44}-P&=\mu\\
  C_{12}+P&= \lambda.
\end{align*}

% Stability imposes that the
% energy is positive definite so that
% \begin{align*}
%   3 \lambda + 2 \mu +\nu >0\\
%   \mu >0\\
%   2\mu+\nu >0
% \end{align*}

The Green function of the static elastic problem is then the inverse
of the Christoffel matrix,
\begin{equation}
  \label{eq:inverse}
  D_{ij}(\bfk) G_{jk}(\bfk) = \delta_{ij}.
\end{equation}%
One expresses the free energy in terms of the displacement field
\begin{equation}%
  F[\hatbfu] = \frac{1}{2} \sum_{\bfk} \hatu_i(\bfk)D_{ij}(\bfk)\hatu_j(\bf-k)\text{.}
\end{equation}
For each wavevector~$\bfk$, $D$ is a three-by-three matrix with
eigenvalues $d_i(\bfk)$ where the subscript $i$ indicates a
polarization state. Following a convention usual in the 
literature~\cite{chaikin,maret2,antinaPRL}, we define the auxiliary
variable~$\omega^2_i(\bfk) = d_i(\bfk)$.

In our previous work we showed that it was useful to perform an
angular average of the elastic constants in order to generate the
``best'' isotropic approximation to the elastic
properties~\cite{Norris06,Fedorov68}. This average gives the
effective Lam\'e constants
\begin{equation}
  \label{eq:fedorov}
  \tillambda = \lambda + \frac{\nu}{5} \quad\text{and}\quad
  \tilmu = \mu + \frac{\nu}{5}.
\end{equation}

On observing a slice of the three-dimensional solid under a confocal
microscope one can then show that the effective dispersion is described
by an anomalous relation where $\omega\sim q^{1/2}$, for small
$\bfq$~\cite{claire}. Here $\bfq$ is the two-dimensional wavevector,
and is to be distinguished from the full three-dimensional wavevector
$\bfk$. A detailed calculation gives
\begin{equation}
  \begin{aligned}
    \label{eq:dispersion}
    \omega_\perp^2 &= 2\tilmu\, q\quad\text{(transverse)}, \\
    \omega_\parallel^2 &=
    \frac{4\tilmu(\tillambda{+}2\tilmu)}{\tillambda+3\tilmu}\,
    q\quad\text{(longitudinal).}
  \end{aligned}
\end{equation}
Thus observation of the two branches of the projected dispersion curve
gives information on the averaged Lam\'e moduli. It is interesting to
note that even in the incompressible limit of rubber elasticity where
$\mu/\lambda\rightarrow 0$ the effective projected frequencies remain
finite. Apparent compression in the two-dimensional plane remains
easy. In this limit we find that
$\omega_\perp^2/\omega_\parallel^2=1/2$.

\section{Discretization}
\label{sec:method}

Rather than performing a molecular dynamics simulation of hard
spheres, we here use only methods from linear algebra to numerically
project and characterize fluctuations from an elastic solid. The
disadvantage is that all non-linearities in the true physical system
are neglected. The advantage is that very efficient methods are
available that generate useful data in just a few minutes of
calculation. One is also able to systematically vary the parameters of
the model in order to study their influence on the projected
fluctuations.

We use a finite difference discretization of
Eq.~\eqref{eq:christoffel}, taking the mesh size as unity and
embedding the elastic medium in a cubic box of dimensions~$L$. For any
wavevector~$\bfk$ we define the components of a vector~$\bfv$ as
\begin{equation}
  v_j= (1 - e^{i k_j}).
\end{equation}
The finite different discretization corresponding to
Eq.~\eqref{eq:christoffel} is then
\begin{equation}
  \label{eq:disc}
  D = \mu\, z(\bfk)\; {\bf I} + (\lambda{+}\mu) (\bfv\otimes\bfv^*) +
  \nu\, \diag(v_i v_i^*),
\end{equation}
with $z(\bfk)= \operatorname{tr}(\bfv\otimes\bfv^*) = \sum_i (2 - 2
\cos{k_i})$ and where ``$\diag$'' constructs a three-by-three matrix
with the given components on the diagonal. ${\bf I}$ is the
three-dimensional unit matrix, and $\otimes$~denotes the exterior
product of two vectors. The matrix~$D$ is complex Hermitian. One sees
that the small-wavevector expansions of Eq.~\eqref{eq:disc} is indeed
given by the Christoffel matrix. The exact analytic form makes the
dispersion relation periodic within the first Brillouin zone of the
discretizing mesh. Note this form of the discretization suffices to
study cuts in the plane $(1,0,0)$. Other cuts are found by rotating
the crystal with respect to the cubic cell, modifying the contribution
in~$\nu$. This is easily performed by updating the tensor~$S$ using
rotation matrices parameterized with Euler angles. In this case the
vectors ${\bf e}^p$ are the three rows of the Euler rotation matrix
and we construct the anisotropic elastic contribution from $ |{\bf e}^p . {\bf v}|^2
e^p_i e^p_j$.

\begin{figure}[tb]
  \centering
  \includegraphics{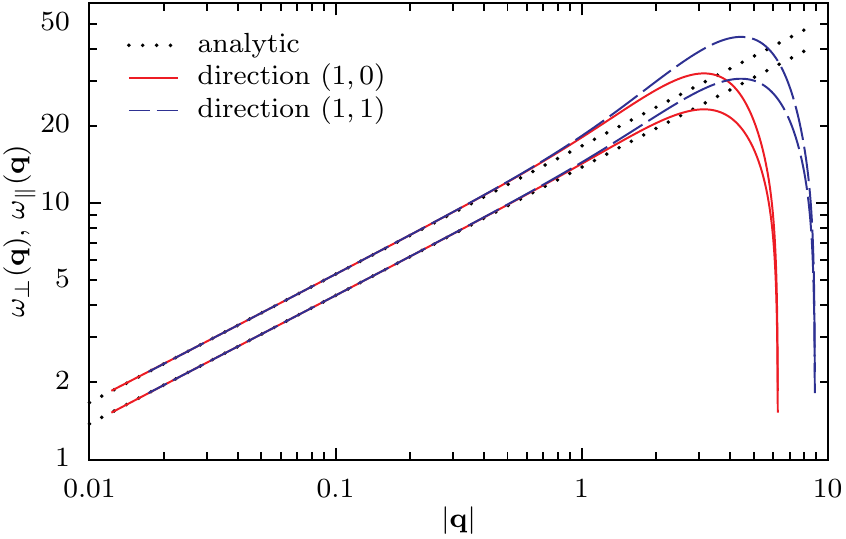}
  \caption{Longitudinal and transverse fluctuations of an isotropic
  system projected to the two-dimensional plane~$(1,0,0)$. The result
  is compared to the long-wavelength limit (dotted line) predicted in
  Eq.~\eqref{eq:dispersion}. $L=512$, $\lambda=70$, $\mu=95$, $\nu=0$.}
  \label{fig:proj}
\end{figure}

We evaluate $D(\bfk)$ for wavevectors $k_i= 2\pi n_i/L$. We
numerically invert the three-by-three matrix $D(\bfk)$ for each value
of $\bfk$ and then use the fast Fourier transform to evaluate the
components of the real space Green function on a cubic grid. We then
evaluate the Green functions on a slice and use two-dimensional fast
Fourier transforms to obtain the effective dispersion relation for a
projected wavevector~$\bfq$. Such a dispersion relation is shown in
Fig.~\ref{fig:proj} for the isotropic case $\nu=0$, where we compare
with the continuum expressions Eq.~\eqref{eq:dispersion}. As expected,
we see agreement between theory and simulation.
We use values of~$\mu$ and~$\lambda$ comparable to those found in our
previous molecular dynamics simulations. The anisotropy~$\nu$ will be
varied over a wide range of values in the figures below. We use units
such that the energy scale is $k_BT$.

We now show how to calculate the full dispersion curve corresponding
to the present discretization, using a alternative calculational route
than that proposed in our original paper.

\subsection{Projection for $\nu=0$}

In our previous work we gave a treatment of the projection which
required calculation of the explicit form of the real-space Green
function. We show here that a  different method allows one to 
calculate the projection for~$\nu=0$, and we apply this method to the
discretized Eq.~\eqref{eq:disc}, generating the analytic expression
for the full curves of Fig.~\ref{fig:proj}. We firstly calculate the
inverse of Eq.~\eqref{eq:disc}
\begin{equation}
  G(\bfk)= \frac{1}{\mu z(\bfk)}\left ({\bf I} -
  \frac{\lambda+\mu}{\lambda+2\mu} \frac{\bfv\otimes\bfv^*}{z(\bfk)}
  \right).
\end{equation}
We now formally calculate the real-space Green function
by Fourier transformation, and back transform the projected
correlations. Elementary calculations show that
\begin{equation}
  G_2(\bfq) = \sum_{k_z} G(q_x, q_y, k_z) \rightarrow
  \frac{1}{2\pi}\int_{-\pi} ^{\pi}
  dk_z  G(\bfq, k_z),
\end{equation}
where $G_2$ describes the correlations projected on the confocal
plane~$z=0$.

The integral is calculated by noting that
\begin{align*}
\frac{1}{2\pi}\int_{-\pi}^{\pi}\frac{1}{A -2\cos{k_z} } dk_z = \frac{1}{(A^2 -4)^{1/2}},\\
\frac{1}{2\pi}\int_{-\pi}^{\pi}\frac{1}{(A -2\cos{k_z})^2 } dk_z = \frac{ A }{(A^2 -4)^{3/2}}.
\end{align*}
Within the plane we consider without loss of generality the choice
$\bfq=(q,0)$, then $A=4-2\cos{q}$. The dispersion relations are then
\begin{align}
  \label{eq:fulldisp}
  \frac{1}{\omega_\perp^2} &=\frac{1}{\mu}
  \frac{1}{\bigl[(4-2\cos{q})^2 -4 \bigr]^{1/2}}\:,
  \\
  \nonumber
  \frac{1}{\omega_\parallel^2} &= \frac{1}{\omega_\perp^2} -
  \frac{1}{\mu}\frac{\lambda+\mu}{\lambda+2\mu}
  \left(
    \frac{(4 - 2\cos{q})(2-2\cos{q}) }{\bigl[(4-2\cos{q})^2 -4\bigr]^{3/2}}
  \right).
\end{align}
When one expands Eq.~\eqref{eq:fulldisp} about $q=0$ one finds exactly
Eq.~\eqref{eq:dispersion}. It should be noted that all higher order
corrections in the expansion of Eq.~\eqref{eq:fulldisp} are
non-universal, and thus change on using different discretizations of
the continuum equations. However these equations are useful in
calibrating the convergence of the numerical code that we developed
for this paper.

\section{Three-dimensional structure of fluctuations}

In this section we discuss the three-dimensional nature of the
fluctuation field induced in the presence of a fluctuation in a
surface. In particular we wish to argue that the use of periodic
boundary conditions does not lead to strong artifacts in the
amplitudes of the measured waves, at the same time we are able to
estimate the decay of boundary perturbations to a sample on
fluctuations within a sample. Very similar arguments are regularly
used in electrostatics \cite{slab,lucas}. Let use firstly summarize the situation for
solutions of the Laplace equation where the algebra is particularly simple.

Consider a periodic box of lateral dimensions $L$ and height $L_z$.
Let there be charges on the plane $z=0$ modulated with wavevector
$k_x$. Away from the plane, in the absence of sources, the potential is
given by a solution to the equation $\nabla^2\phi=0$. If the solution
is periodic in the $x-y$ plane then the solution must decay
exponentially in the $z$ direction so that
\begin{equation}
\phi = \phi_k\exp{(ik x - k |z|)}.
\end{equation}
The slowest decaying wave in the $z$ direction has $k=2\pi/L$, thus successive images in the
box have an asymptotic interaction \cite{slab} that varies as
\begin{math}
 \sim e^{-2 \pi L_z/L}.
\end{math}
Already in a cubic box, for which $L_z=L$ the image interaction via
this component of the density is strongly suppressed. Higher order
waves have even weaker interactions.

We now return to the problem of an isotropic elastic solid.
If we observe a sinusoidal perturbation on the plane $z=0$ we can find
the minimum energy fluctuation consistent with this observation by
examining the solution of the equation
\begin{equation}
(1-2 \sigma) \nabla^2 {\bf u} + \grad \div {\bf u}=0
\end{equation}
away from the plane $z=0$, where $\sigma$ is the Poisson ratio of the
material. We  look for solutions of the form
\begin{equation}
  {\bf u} = e^{i k x} (u(z), 0, w(z) )
\end{equation}
% \begin{equation}
% \div \rightarrow e^{ikx} (iku+ w_z )
% \end{equation}
% \begin{equation}
% \grad \div \rightarrow ( ik (iku+w_z) ,0, iku_z+w_{zz} )
% \end{equation}
% \begin{equation}
% \nabla^2 \rightarrow (-k^2 u + u_{zz} ,0, -k^2w + w_{zz})
% \end{equation}
and find
\begin{align*}
  (1-2\sigma)( -k^2 + \gamma^2)u-k^2 u + i k \gamma w &=0,\\
  (1-2\sigma)( -k^2 + \gamma^2)w +\gamma^2 w+ik\gamma u &=0,
\end{align*}
where $\gamma$ is the Laplace transform of the fields in the $z$ direction.
Solutions are found when the determinant of coefficients is zero which gives
\begin{equation*}
  %(-k^2 + \epsilon(-k^2+\lambda^2)(\lambda^2 + \epsilon(-k^2
  %+\lambda^2)) +k^2\lambda^2&=0\\
  k^2 = \gamma^2,
\end{equation*}
implying a result very similar to that in electrostatics: There is a
decay of interactions on a scale which is determined by the wavelength
of the observed mode, independent of the Poisson ratio. 
There is also  a simple phase relationship between the horizontal and
vertical components of the fluctuations:
\begin{equation*}
  u\pm iw=0
\end{equation*}

From this calculation we understand that we should only observe
samples far from any external walls. Fluctuations of wavelength $\ell$
observed within a plane decay over a distance $\ell/2\pi$ and will be
perturbed by external surfaces which are too close.  This result
allows us to find a simple physical interpretation of the anomalous
dispersion relation: Imposition of a fluctuation of wavelength $\ell$
leads to an energy density $(u/\ell)^2$. This excites a total volume
of the sample which varies as $L^2\ell$, giving a scaling in the total
elastic energy of mode~$\bfq$ as $F_\bfq \sim 1/\ell \sim |q|$.
%  We
%also note that in electrostatics the mode $q=0$ can lead to
%%exceptional properties in the energy, we do not pursue the point for
%elasticity, where it does not influence the analysis of modes where $
%q \ne 0$.

We now study the variation of the elastic properties  as a function
of the cubic anisotropy $\nu$, and compare the results with the
averaging approximation.
\begin{figure}[tbh]
  \centering
  \includegraphics{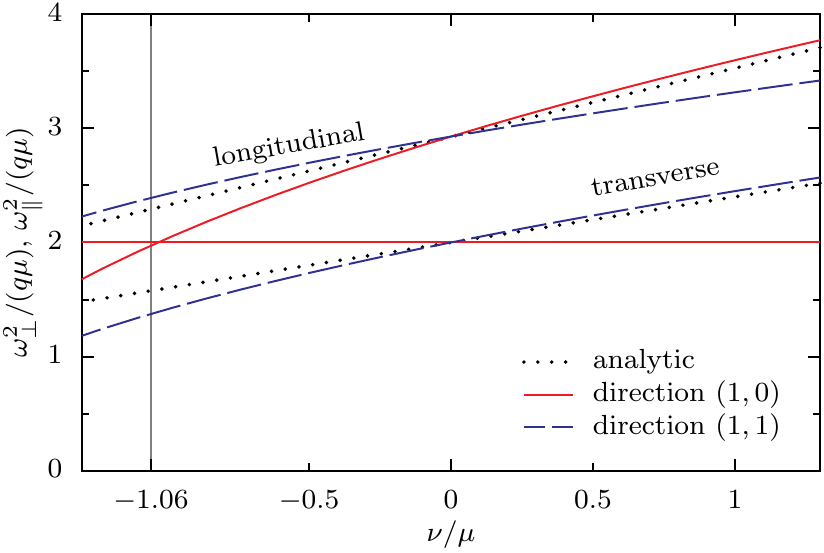}%
  \caption{Evolution of amplitudes as a function of $\nu$ in a cut
  perpendicular to $(1,0,0)$. Previous theory in dotted line,
  Eq.~\eqref{eq:dispersion}. The theoretical curves are intermediate
  between the two numerically generated curves. The curve in the
  direction $(1,0)$ displays the exceptional features to be
  independent of $\nu$. All lines intersect at $\nu=0$ since the
  system becomes isotropic. $L=512$. $\lambda/\mu=0.73$.}%
  \label{fig:100}%
\end{figure}

\section{Numerical Results}

We implemented the Fourier and matrix manipulation using an
Octave/Matlab script; the language was chosen for its facilities for
the manipulation of fast Fourier transforms and matrix algebra. We
projected the fluctuations onto different planes of the crystal and
studied the effective dispersion relations, Fig.~\ref{fig:100}. For the
cut of the crystal in the plane $(1,0,0)$ we plot the effective
dispersion relations in two directions in the plane, $(1,0)$ and
$(1,1)$. The averaging approximation, while not wholly inaccurate,
does not account for a number of qualitative features of the measured
dispersion curves. In particular, the measured transverse stiffness in
the direction $(1,1)$ is \emph{independent} of~$\nu$ while the
longitudinal mode varies strongly with the anisotropy. This gives rise
to  level crossing for large negative values of~$\nu$, which occurs
for values of the parameters very close to those found for the hard
sphere system.

\begin{figure}[htb]
  \centering
  \includegraphics{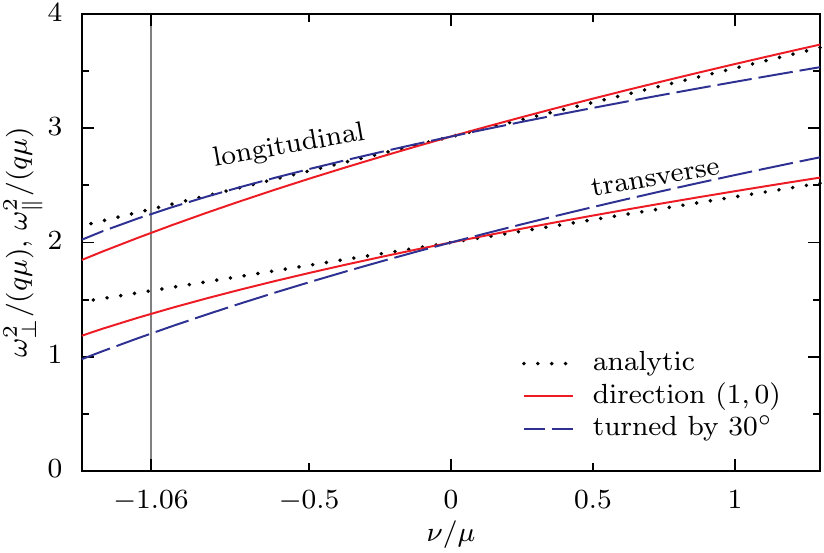}
  \caption{Evolution of amplitudes as a function of $\nu$
     in a cut perpendicular to $(1,1,0)$. $L=512$. $\lambda/\mu=0.73$.}
  \label{fig:110}
\end{figure}

The situation in the plane $(1,1,0)$, Fig.~\ref{fig:110}, is somewhat
closer to the approximate analytic form of the dispersion curves. Here
again the dispersion is anisotropic within the plane. Clearly this is
impossible to describe with spherically averaged elastic constants.

\begin{figure}[htb]
  \centering
  \includegraphics{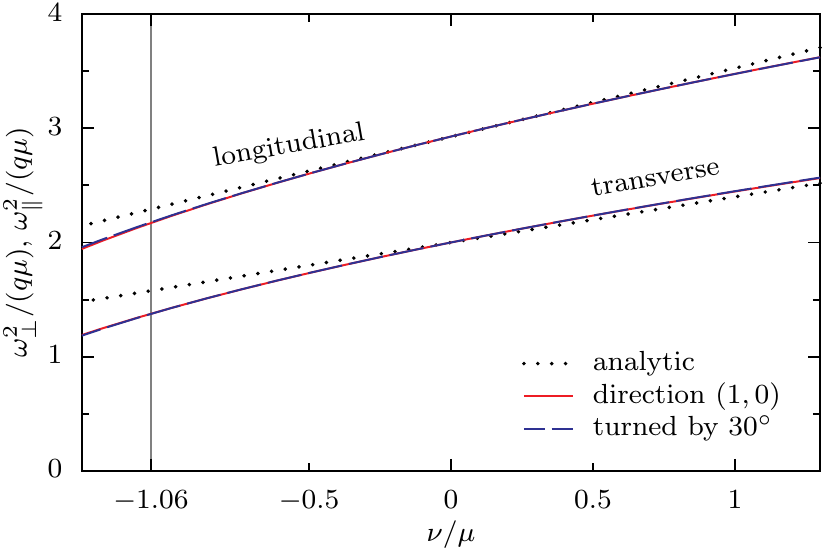}
  \caption{Evolution of amplitudes as a function of $\nu$ in a cut
  perpendicular to $(1,1,1)$. The dispersion curve is isotropic in the
  plane. $L=512$. $\lambda/\mu=0.73$.}
  \label{fig:111}
\end{figure}

For the plane $(1,1,1)$, Fig.~\ref{fig:111} we see that the averaging
procedure works particularly well and that the dispersion relations
are isotropic. This is the curve which is the most useful
experimentally, because the samples naturally grow from flat surfaces
aligned in this orientation.

% \begin{figure}[htb]
%   \centering
%   \includegraphics[width=8cm]{ee}
%   \caption{$L=128$. Evolution of longitudinal amplitude as a function
%      of $\lambda$ and $\nu$ on the plane  $(1,1,1)$.}
%   \label{fig:ee}
%\end{figure}

The vertical gray lines in Figs.~\ref{fig:100}--\ref{fig:111} at
$\nu/\mu=-1.06$ represent the anisotropy found in the full
molecular-dynamics simulation. Also the ratio~$\lambda/\mu$ has been
chosen to fit to the simulation. Comparing the published
data~\cite{claire} with Fig.~\ref{fig:111}, we notice that the full
simulation curves are displaced with respect to the data produced by
Fourier analysis. Both are slightly displaced with respect to the
isotropic average, but with different signs. We interpreted this shift
as being due to non-linearities in the full system and estimated it to
about $3\%$. A new estimate based on the numerical method described in
this paper is larger, about $9\%$ and~$15\%$.

\subsection{Analytic projection on $(1,0,0)$}

Using the method of Sec.~\ref{sec:method}, it is easy to write a
formal solution for the projected properties for arbitrary~$\nu$. In
the limit of small~$q$ one finds as integrand the ratio of two
polynomials of orders four and six in the vector~$(q_x, q_y, k_z)$,
see also~\cite{Morawiec94}. In principle by factorizing the
denominator (which is of order six) one has the full analytic solution
to the problem. In practice this requires the roots of high order
equations and leads to clumsy and opaque expressions. There is,
however, one case which is particularly easy to understand, namely the
independence of the transverse mode in Fig.~\ref{fig:110} on~$\nu$.
This is a simple consequence of the Sherman-Morrison identity: The
Christoffel matrix on the plane $(1,0,0)$ can be written as a diagonal
matrix plus a rank one update:
\begin{align*}
  D &= B + (\lambda+\mu) ( \bfk\otimes\bfk),\\
  \text{with}\quad B&=\diag( \mu k^2 +\nu k_i^2).
\end{align*}
The inverse of this matrix is then
\begin{equation}
  G= B^{-1} - \frac{(\lambda+\mu)(B^{-1}\bfk) \otimes
  (B^{-1}\bfk)}{1+(\lambda+\mu)\bfk B^{-1}\bfk}.
\end{equation}
If $\bfq=(1,0)$, then it is clear that $G_{xy}(1,0,k_z)=0$ and the
only contribution to the projected transverse spectrum comes from
$G_{yy}(1,0,k_z)$ which is trivially independent of~$\nu$.

The longitudinal contribution can also be found from the
Sherman-Morrison formula, and we find the integral:
\begin{multline*}
 \frac{ \mu q}{\omega^2_\parallel}  = \int_{-\pi}^{\pi}
 \frac{dk}{2\pi}  \frac{1}{ 1+ \nu/\mu + k^2}\:\times \\
 \left [ 
1- 
\frac{   1+ k^2+k^2 \nu/\mu      }
{\frac{ (1+\nu/\mu +k^2)(1+k^2+ k^2\nu/\mu)}{1+\lambda/\mu}
      + 1+ 2 (1+\nu/\mu)k^2 +k^4  } \right ] 
\end{multline*}
This can be integrated by finding the roots of the denominator, and
expanding using partial fractions, the result is however too
complicated to be illuminating.

%den[z_]= (1+z^2)(1+n+z^2)/l + (1+z^2(1+n+z^2))
%f[z_] = ((1+z^2)/(1+n+z^2)) * 1/den[z];
%g=Integrate[ f[z], {z,-Infinity, Infinity},Assumptions ->
%n \[Element] Reals && l \[Element] Reals ]

\subsection{Simulation data}

\begin{figure}[htb]
  \centering
  \includegraphics{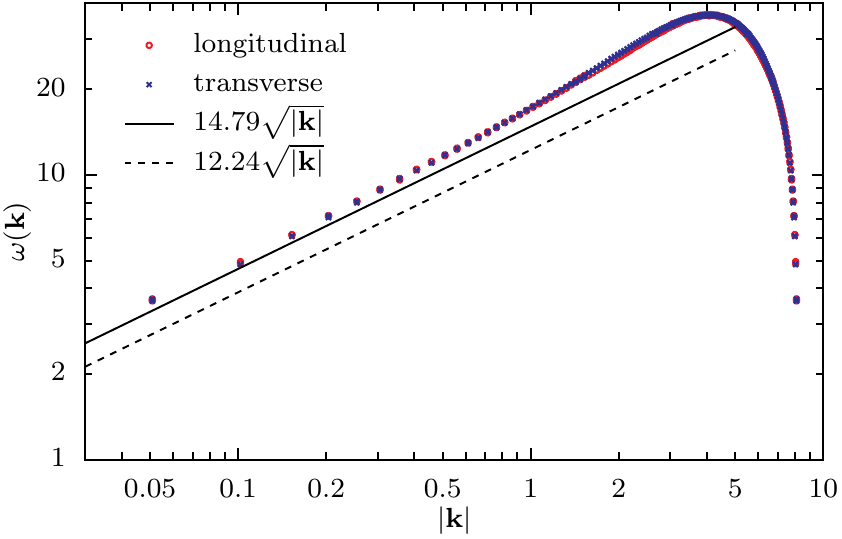}%
  \caption{Effective dispersion relations for the plane $(1,0,0)$ in
  the direction $(1,0)$. Dataset of~\cite{claire}. The longitudinal
  and transverse branches are almost degenerate for large negative
  cubic anisotropy.}%
  \label{fig:hann}%
\end{figure}

On seeing the surprising nature of the dispersion in
Fig.~\ref{fig:100} where the longitudinal and transverse modes become
almost degenerate at large negative values of the anisotropy, around
$\nu\sim-\mu$, we performed an analysis based on our molecular
dynamics simulation, Fig.~\ref{fig:hann}. There, the anisotropy was
also strong, $\nu=-1.06\mu$. We find that indeed our numerical data
displays very similar features, confirming the useful nature of the
Fourier based code for exploring the properties of the elastic
behavior of hard sphere systems. The longitudinal and transverse
dispersion relations are almost degenerate over all wavevectors in the
direction $(1,0)$. We do not have any simple explanation for this
result.

\begin{figure}[htb]
  \centering
  \includegraphics{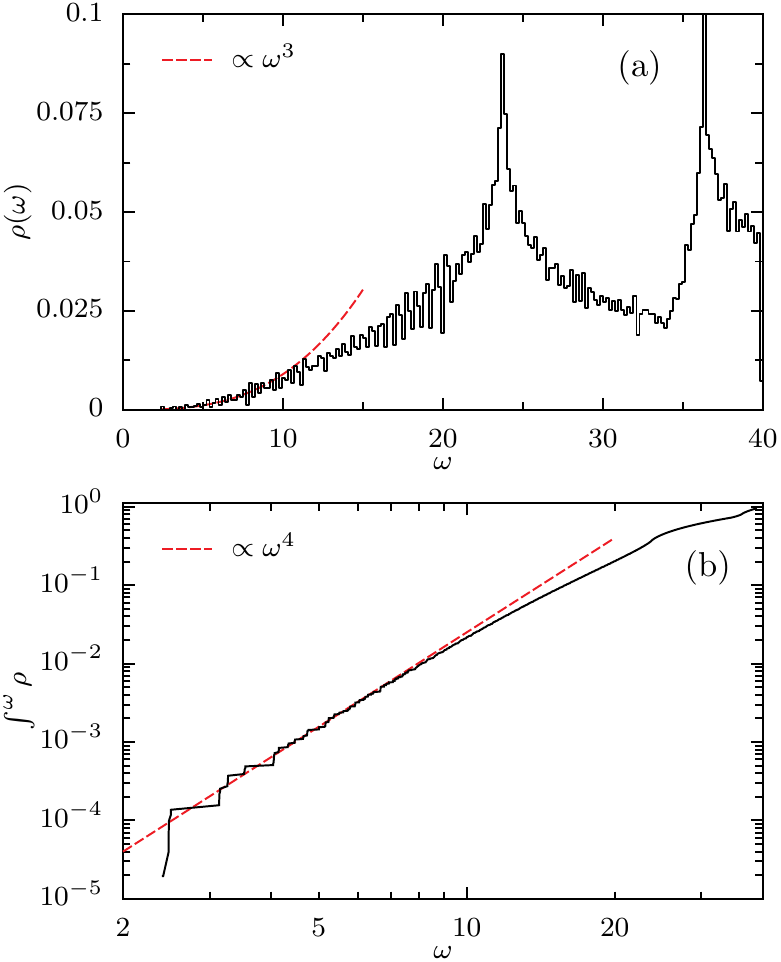}
  \caption{Density of states measured on the $(1,1,1)$ cut, data from
  the molecular dynamics simulation~\cite{claire}. (a)~Note the
  prominent van Hove singularities corresponding to the longitudinal
  and transverse projected modes. Panel~(b) plots the integrated
  density of states on a log-log scale, compared with the theoretical
  form $\omega^4$.}
  \label{fig:states}
\end{figure}
\begin{figure}[htb]
  \centering
  \includegraphics{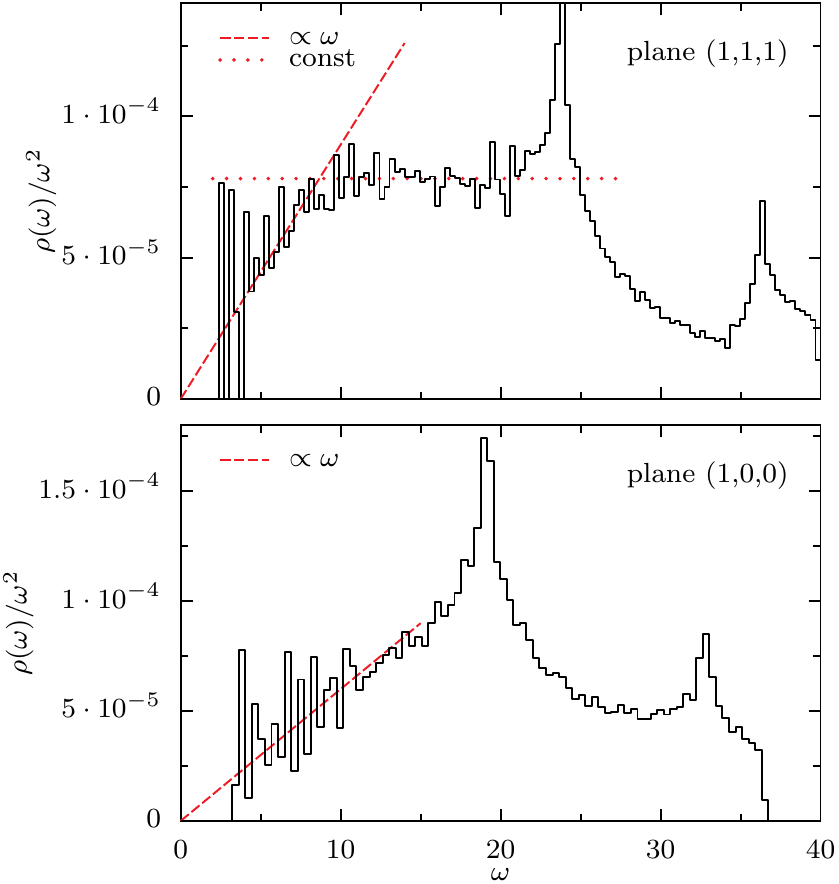}
  \caption{Density of states divided by~$\omega^2$, projected on two
    different planes. The resulting curves depend on the projection
    plane: In the case $(1,1,1)$ the expected regime in $\rho\sim
    \omega^3$ is seen for $\omega<10$; a plateau is observed for
    $10<\omega<20$. The form $\rho\sim \omega^3$ fits over a larger
    range of $\omega $ for the case $(1,0,0)$.}
  \label{fig:plateau}
\end{figure}

\section{Density of states}

In the interpretation of the density of states of disordered material
one often compares with the Debye theory of elasticity. This allows
one to establish an expectation for the low-frequency mode structure,
and then to determine whether the disorder has given rise to a deficit
or excess in the density of states. As noted in~\cite{ghosh} a
dispersion relation such as Eq.~\eqref{eq:dispersion} leads to
modifications in the Debye spectrum, which is normally given as
$\rho(\omega)\sim\omega^{d-1}$ for a \hbox{$d$-dimen}\-sional elastic solid. If
we use instead the dispersion relations in Eq.~\eqref{eq:dispersion}
and define $\omega_\parallel^2 = a_\parallel q$ and $\omega_\perp^2 =
a_\perp q$ we find that the density of states, expressed in terms of
the variable~$\omega$ is given by
\begin{equation}
  \rho(\omega) = \frac{1}{\pi} \Bigl(\frac{1}{a_\parallel^2} +
  \frac{1}{a_\perp^2}\Bigr) \omega^3,
\end{equation}
which corresponds to neither the projected, nor the embedding
dimension of the slice. Figure~\ref{fig:states} shows the density of
states of a cut of our molecular dynamics crystal. The top panel
displays the binned density of states on a linear scale. In order to
avoid artifacts of binning we also plot the integrated density of
states in the bottom panel and find that there is a good fit to a law
in~$\omega^4$ for $\omega<8$. Some of the experimental literature
plots two-dimensional slices normalized by the Debye density of states
for a three-dimensional crystal~\cite{science}. We have also plotted
this form in the top panel of Fig.~\ref{fig:plateau}. We see, rather
surprisingly, a region between the local frequency behavior in
$\omega^3$ and the first peak where the normalized data seems
relatively constant. This seems, however, to be a simple cross-over
occurring over a modest range of frequencies. If we project on a
different plane, $(1,0,0)$ in the bottom panel of
Fig.~\ref{fig:plateau}, no plateau is visible.

% From Landau and Lifshitz Eq. (7.4)
% \begin{equation}
% G_{ik} = \frac{1+\sigma} {8 \pi E (1-\sigma)} 
% \left [ (3-4 \sigma)
%   \delta_{ik} +n_i n_k \right ] \frac{1}{r}
% \end{equation}
% $E=2\mu(1+\sigma) $
% this looks quite regular near $\sigma=1/2$.

% The fourier transform to two dimensions gives
% \begin{equation*}
% G= \frac{1}{q} (2- \hat \bfq \hat \bfq)
% \end{equation*}

% possible explanations-

\section{Conclusions}
We have introduced a Fourier based method which can efficiently
generate the Green function in anisotropic elastic media. It allows
one to rapidly perform detailed studies of the evolution of correlations
and fluctuations in projected geometries, such as those used in
several recent experiments. We have studied the effective dispersion
relations of a colloidal crystal projected on the plane $(1,0,0)$ and
shown that the near degeneracy of the dispersion curves in the
direction $(1,0)$ is due to the large negative cubic anisotropy.
Surprisingly long wavelengths are needed to see the continuum limit in the
density of states. Studies on certain cuts lead to an intermediate
regime for the density of states.

Code written in Octave/Matlab is available as supplementary material
to the paper to perform the Fourier based evaluation of Green
functions, together with the projection to two dimensions.

%\bibliographystyle{epj}
%\bibliography{paper}

\end{document}